\documentclass[aps,prd,twocolumn,showpacs,nofootinbib]{revtex4}
\pdfoutput=1

\usepackage{graphicx}
\usepackage{bm}
\usepackage{amssymb,amsmath,latexsym}
\usepackage{color}
\usepackage{bbold}
\usepackage{physymb}
\usepackage[colorlinks=true,linktocpage=true,linkcolor=blue,citecolor=blue]{hyperref}

\newcommand{\beq}{\begin{equation}}
\newcommand{\eeq}{\end{equation}}
\newcommand{\bqa}{\begin{eqnarray}} 
\newcommand{\eqa}{\end{eqnarray}}

\newcommand{\rhot}{\tilde{\rho}}

\newcommand{\J}{\tilde{{\cal J}}}

\newcommand{\pit}{\tilde{\pi}}
\newcommand{\Pit}{\tilde{\Pi}}

\newcommand{\dft}{\delta\tilde{f}}

\newcommand{\mud}{\mu_1\cdots\mu_\ell}

\newcommand{\bp}{{\bm{p}}}
\newcommand{\bk}{{\bm{k}}}
\newcommand{\bK}{{\bm{K}}}
\newcommand{\br}{{\bm{r}}}

\newcommand{\tf}{{\tilde{f}}}
\newcommand{\eq}{{\,=\,}}

\newcommand{\rs}{{\rm a}}

\begin{document}


\title{Transient oscillations in a macroscopic effective theory of the Boltzmann equation}

\author{Dennis Bazow}
\author{Mauricio Martinez}
\author{Ulrich Heinz}
\affiliation{Department of Physics, The Ohio State University,
  Columbus, OH 43210, United States}

\begin{abstract}
A new transient effective theory of the relativistic Boltzmann equation is derived for locally momentum-anisotropic systems. In the expansion of the distribution function around a local ``quasi-equilibrium'' state a non-hydrodynamic dynamical degree of freedom is introduced at leading order that breaks local momentum isotropy. By replacing the deviation of the distribution function from this quasi-equilibrium state in terms of moments of the leading-order distribution and applying a systematic power counting scheme that orders the non-hydrodynamic modes by their microscopic time scales, a closed set of equations for the dynamical degrees of freedom is obtained. Truncating this set at the level of the slowest non-hydroynamic mode we find that it exhibits transient oscillatory behavior -- a phenomenon previously found only in strongly coupled theories, where it appears to be generic. In weakly coupled systems described by the Boltzmann equation, these transient oscillations depend on the breaking of local momentum isotropy being treated non-perturbatively at leading order in the expansion of the distribution function.
\end{abstract}

\date{\today}

\maketitle


\section{Introduction}
The Boltzmann equation provides a microscopic link between statistical physics and macroscopic fluid dynamics. For weakly coupled, dilute plasmas of massive particles for which two-particle correlations can be neglected the Boltzmann equation with only elastic binary collisions yields a good description of the microscopic physics. While hydrodynamics provides an effective theory for the evolution of conserved macroscopic quantities (energy, momentum, and conserved charges), a full description of the microscopic dynamics of the Boltzmann equation must include an infinite set of non-hydrodynamic modes. Hydrodynamic modes have dispersion relations that satisfy $\lim_{\bk\to 0}\omega_{n}(\bk)=0$; all other modes are called non-hydrodynamic. The former dominate the dynamics of the system at large spatial and temporal length scales. The influence of non-hydrodynamic modes is suppressed for small values of the Knudsen number Kn (the ratio between the mean free path and the typical macroscopic length scale of the system). 

Systematic coarse-graining leads to effective theories with varying degrees of freedom and non-hydrodynamic modes. For relativistic systems, causal effective theories must include at least the first non-hydrodynamic mode in the dynamics of the dissipative currents \cite{Denicol:2011fa, Noronha:2011fi}. Starting from the Boltzmann equation and expanding the distribution function around an isotropic local equilibrium state it was shown \cite{Denicol:2011fa} that all non-hydrodynamic modes have frequencies that for $\bk{\,\to\,}0$ lie on the negative imaginary axis of the complex $\omega$ plane. At large times, only the slowest non-hydrodynamic mode with the smallest frequency $|\omega|$ survives. This, together with a power counting scheme in Knudsen number Kn and inverse Reynolds numbers $\mathrm{R}^{-1}_{i}$ (the ratios between the dissipative currents and the thermal equilibrium quantities), allowed Denicol{\it~et\,al.} to systematically derive a resummed transient relativistic fluid dynamical theory \cite{Denicol:2012cn}. 

The approach in \cite{Denicol:2012cn} is constrained by expanding the Boltzmann equation  around an isotropic local equilibrium distribution, and hence by treating all deviations from local momentum anisotropy perturbatively. The infinite moment hierarchy is truncated by a power counting scheme based on inverse Reynolds numbers that include all dynamically generated momentum anisotropies. When a system expands strongly anisotropically, such as the quark-gluon plasma fireball created during the early stages of a relativistic heavy-ion collision \cite{Strickland:2014pga,Heinz:2014zha}, these local momentum anisotropies can become large and must be treated nonperturbatively. Here we do so by modifying the local thermal equilibrium distribution by introducing a non-hydrodynamic degree of freedom describing the deviation from momentum isotropy, thereby defining an anisotropic quasi-equilibrium state \cite{fn1}. In a procedure that closely follows \cite{Denicol:2012cn}, but now resums and truncates the moment hierarchy according to a modified power counting scheme based on the Knudsen and {\em residual} Reynolds numbers from which the largest contributions arising from the local momentum anisotropy have been eliminated \cite{reynolds}, we derive a new type of transient relativistic fluid dynamics in which the slowest non-hydrodynamic mode (associated with the local momentum anisotropy) turns out to exhibit transient oscillations.
\\
\indent
Such oscillations appear to be generic in macroscopic effective theories of strongly coupled plasmas \cite{Starinets:2002br, Nunez:2003eq, Chesler:2008hg, Noronha:2011fi, 
Heller:2014wfa}, but were thought not to exist in analogous theories of weakly coupled systems that are based on the Boltzmann equation \cite{Denicol:2011fa,Noronha:2011fi}. We find that they arise generically also in weakly coupled systems and are absent only if the macroscopic effective theory is based on a perturbative expansion around an isotropic local equilibrium state.


\section {Derivation} We start from the relativistic Boltzmann equation for a gas of particles with mass $m$, 
\begin{equation} 
\label{BE}
  k^{\mu }\partial _{\mu }f(x,k)=C[f](x,k),
\end{equation}
where $k^\mu\eq(E_k,\bk)$, with $E_k\eq\sqrt{\bk^2{+}m^2}$, is the particle four-momentum. We write shortly $f_\bk{\,\equiv\,}f(x,k)$, $C_\bk \equiv C(x,k)$, and use $dK=d^3k/[(2\pi^3)E_k]$ to denote the Lorentz invariant momentum integration measure. The collision term for elastic 2-body collisions is \cite{fn2a}
\begin{eqnarray}
\label{colKern}
 C_\bk[f] &=&\frac{1}{2}\int dK'dPdP'\,W_{\bk\bk'\to \bp\bp'}  
\notag \\
&&\times 
\left( f_\bp f_{\bp'}\tf_\bk \tf_{\bk'}{-}f_\bk f_{\bk'} \tf_\bp \tf_{\bp'} \right),
\end{eqnarray}
where $W_{\bk\bk'\to \bp\bp'}$ is the transition rate for the process $k+k'\to p+p'$ and $\tf\equiv 1-af$, where $a=\pm 1,0$ for fermions, bosons, or classical distinguishable particles, respectively.

We consider systems undergoing highly anisotropic collective expansion, assumed to be stronger in the longitudinal ($z$) than in the transverse ($\br_\perp{\,\equiv\,}(x,y)$) directions. In dissipative systems this leads to a local momentum anisotropy which, for simplicity, we take at leading order to be azimuthally symmetric around $z$ in the local rest frame (LRF), treating non-leading corrections to the momentum anisotropy as small perturbations. These assumptions are implemented by writing \cite{Bazow:2013ifa} $f_\bk{\eq}f_{\rs,\bk}{\,+\,}\delta\tf_\bk \equiv f_{\rs,\bk}\bigl(1{+}\tf_{\rs,\bk}\phi_\bk\bigr)$  where in the LRF the momentum-ani\-sotropic (``a'') leading-order term takes the form \cite{Romatschke:2003ms} 
\beq
\label{fa}
f_{\rs,\bk} = f_\mathrm{eq}
\left(\beta_\rs\sqrt{m^2{+}k^2_\perp{+}(1{+}\xi)k^2_z}\right),
\eeq
with the equilibrium distribution $f_\mathrm{eq}(\zeta)\eq1/(e^\zeta{+}a)$. The deviation $\phi_\bk$ is assumed to be ${\,\ll\,}1$. $\xi(x)$ parametrizes the leading-order momentum anisotropy at point $x$, and $\beta_\rs(x)$ is the local ``transverse temperature'' at that point.

The dynamics of the Boltzmann equation is characterized by a hierarchy of microscopic time scales. The temperature parameter $\beta_\rs$, related to the energy density at point $x$ whose evolution is controlled by energy conservation, varies on hydrodynamic time scales. The evolution of the anisotropy parameter $\xi$, like that of all other dissipative flows, is not constrained by conservation laws and thus happens on the microscopic time scales associated with the Boltzmann collision term. We are interested in the ``transient anisotropic regime'' where the fast microscopic processes have all decayed except for the ones associated with the slowest of these microscopic time scales.

The conserved particle current and energy-momentum tensor are given by the first and second moments of the distribution function (with $\left\langle\cdots\right\rangle{\,\equiv\,}\int dK\left( \cdots \right) f_\bk$):
\beq
\label{currents}
J^{\mu } =\left\langle k^\mu\right\rangle, \quad
T^{\mu \nu } =\left\langle k^{\mu }k^{\nu }\right\rangle.
\eeq
The velocity field $u^{\mu}$ (Landau frame) is defined as the timelike eigenvector of $T^{\mu\nu}$:
%
$T^\mu_{\ \,\nu} u^\nu  ={\cal E} u^\mu$. 
%
Its eigenvalue ${\cal E}$ is the LRF energy density. We decompose the four-momenta into temporal and spatial LRF components:
\begin{equation}
\label{decomp}
k^{\mu}= (u{\cdot}k)u^{\mu}+k^{\langle\mu\rangle}
= E_\bK u^{\mu}+\sum_{i=1}^3 K_{i}X^\mu_{i}.
\end{equation}
The four-vectors $X_i^\mu$ reduce in the LRF to unit vectors along the 3 spatial  directions, and $(E_\bK,\bK)$ are the LRF components of $k^\mu$. Using the generalized Landau matching condition that $\beta_\rs$ in (\ref{fa}) be chosen such that $\delta \tf_\bk$ does not contribute to the LRF energy density ${\cal E}$, one finds the anisotropic viscous hydrodynamic decomposition \cite{Bazow:2013ifa}
\begin{eqnarray}
\label{vahydro}
&&\, J^{\mu}={\cal N}u^{\mu}+\tilde{\cal N}^{\mu}\,,
\notag\\
&&\, T^{\mu\nu}={\cal E}u^{\mu}u^{\nu}-({\cal P}_{\perp}+\Pit)\Delta^{\mu\nu}+{\cal P}^{\mu\nu}+\pit^{\mu\nu}\,,\qquad
\\
\label{decomposition}
&&\begin{array}{ll}
{\cal N}\equiv\langle u{\cdot}k\rangle_{\rs},
& \tilde{\cal N}^{\mu}\equiv \langle k^{\langle\mu\rangle}\rangle_{\tilde\delta},
\\
{\cal E}\equiv\langle (u{\cdot}k)^{2}\rangle_{\rs},
& \Pit\equiv-\frac{1}{3}\langle\Delta^{\alpha\beta}k_{\alpha}k_{\beta}\rangle_{\tilde\delta},
\\
\pit^{\mu\nu}\equiv\langle k^{\langle\mu}k^{\nu\rangle}\rangle_{\tilde\delta},
& {\cal P}_{L}\equiv \langle k_z^2\rangle_{\rs},
\\
{\cal P}_{\perp}\equiv\frac{1}{2}\langle (k_x^2+k_y^2)\rangle_{\rs},
& {\cal P}^{\mu\nu}\equiv ({\cal P}_{L}-{\cal P}_\perp)z^{\mu}z^{\nu},
\end{array}
\end{eqnarray}
with $\langle\cdots\rangle_\rs{\,\equiv\,}\int dK(\cdots)f_{\rs,\bk}$,  $\langle\cdots\rangle_{\tilde\delta}{\,\equiv\,}\int dK(\cdots)\dft_\bk$, $z^\mu$ $\equiv X^\mu_3$ (i.e. the $z$ unit vector in the LRF), and $A^{\langle \mu \nu\rangle}\equiv\Delta^{\mu\nu}_{\alpha\beta}A^{\alpha\beta}$ where $\Delta^{\mu\nu}_{\alpha\beta}\equiv\Delta^{(\mu}_\alpha\Delta^{\nu)}_\beta - \Delta^{\mu\nu}\Delta_{\alpha\beta}/3$ and $\Delta^{\mu\nu}\equiv g^{\mu\nu}{-}u^\mu u^\nu= {-}\sum_i X^\mu_iX^\nu_i$. Note that $A^{\langle\mu\rangle}$ defined in (\ref{decomp}) can also be written as $A^{\langle\mu\rangle}\eq\Delta^{\mu\nu}A_{\nu}$.

Following \cite{Denicol:2012cn}, we now replace the Boltzmann equation by a hierarchy of moment equations. We expand the deviation $\phi_\bk(x)$ in a complete orthogonal basis of irreducible tensors $1,\, k^{\langle\mu\rangle},\, k^{\langle\mu}k^{\nu\rangle},\,  k^{\langle\mu}k^{\nu}k^{\lambda\rangle},\, \ldots$, constructed from the spatial components of $k$ in the LRF at point $x$ \cite{DeGroot}:
\begin{equation}
\label{phi}
 \phi_\bk(x)=\!\!\sum_{\ell=0}^{\infty} \Bigl(\sum_{n=0}^{N_\ell}  c^{\langle\mu_1\cdots\mu_\ell\rangle}_{n}(x) P^{(\ell)}_{n}\bigl(x,k{\cdot}u(x)\bigr)\Bigr) k_{\langle\mu_1}\cdots k_{\mu_\ell\rangle}.
\end{equation}
The $P^{(\ell)}_{n}\equiv\sum_{r=0}^{n}a^{(2)}_{nr}(x)\bigl(u(x){\cdot}k\bigr)^{r}$ are orthogonal polynomials of order $n$ in the LRF energy $u{\cdot}k$. Their orthogonality can be used to express the expansion coefficients $c^{\langle\mud\rangle}_{n}(x)$ in terms of the irreducible moments of the residual distribution $\dft_\bk$ \cite{Denicol:2012cn,fn2} 
\begin{equation}
\label{rhoDef}
  \rhot^{\mu_1\cdots\mu_\ell}_{r}(x)\equiv 
  \int dK (u{\cdot}k)^{r}k^{\langle\mu_1}\cdots k^{\mu_\ell\rangle}\dft_\bk\;.
\end{equation}
After some algebra \cite{bazowIP} the distribution function is re\-written as a series in the irreducible moments:
\begin{eqnarray}
\label{momentexpansion}
&&f_\bk=f_{\rs,\bk}\Bigl(1+\tf_{\rs,\bk}\sum_{\ell=0}^{\infty}
\sum_{k=0}^{\infty}\sum_{r=0}^{N_k}\bigl({\cal H}^{(\ell k)}_{\bk r}\bigr)
^{\langle\mu_1\cdots\mu_\ell\rangle}_{\langle\nu_1\cdots\nu_k\rangle}
(x,u{\cdot}k)
\notag\\
&&\qquad\qquad\qquad\qquad
\times\ \tilde{\rho}^{\nu_1\cdots\nu_k}_{r}(x)\,k_{\langle\mu_1}\cdots k_{\mu_\ell\rangle}\Bigr)\;,
\end{eqnarray}
where $\bigl(\tilde{\cal H}^{(\ell k)}_{\bk r}\bigr)^{\langle\mu_1\cdots\mu_\ell\rangle }
_{\langle\nu_1\cdots\nu_k\rangle}{\equiv}\sum_{n=0}^{N_\ell}P^{(\ell)}_{n}(u{\cdot}k)
\bigl[(H^{-1})^{(\ell k)}_{nr}\bigr]^{\langle\mu_1\cdots\mu_\ell\rangle }
_{\langle\nu_1\cdots\nu_k\rangle}$ with 
\begin{eqnarray}
&&\bigl(H^{(\ell k)}_{mr}\bigr)^{\langle\mu_1\cdots\mu_\ell\rangle} _{\langle\nu_1\cdots\nu_k\rangle}(x)\equiv
\\\notag
&&\sum_{s=0}^{m}a^{(k)}_{ms}
\int dK (u{\cdot}k)^{r+s} k^{\langle\mu_1}\cdots k^{\mu_\ell\rangle}
k_{\langle\nu_1}\cdots k_{\nu_k\rangle} f_{\rs,\bk} \tf_{\rs,\bk}.
\end{eqnarray}

From here on we consider for simplicity a gas of massless particles with vanishing chemical potential \cite{bazowIP} for which energy and momentum are the only conserved quantities. Using (\ref{vahydro}), (\ref{decomposition}), and (\ref{fa}), their conservation law $\partial_\mu T^{\mu\nu}\eq0$ then yields the following hydrodynamic equations (see \cite{Bazow:2013ifa} for the definition of the modified thermodynamic $\tilde{\cal J}$-integrals over $f_{\rs,\bk}$):
\begin{eqnarray}
\label{consLaws}
&&\dot{\beta}_\rs=\frac{1}{\J_{2,0,-1}}\left[
\frac{\beta_\rs}{2}\J^{zz}_{2,0,-1}\dot{\xi}+\left({\cal E}+\frac{2{\cal P}_\perp{+}{\cal P}_L}{3}\right)\theta
\right.
\notag\\
&&\hspace{2.3cm}\left .- ({\cal P}^{\mu\nu}{+}\pit^{\mu\nu})\sigma_{\mu\nu}
\frac{}{}\right] \;,
\\\notag
&&\dot{u}^{\mu}=\left[\nabla^{\mu}{\cal P}_\perp-\Delta^{\mu}_{\beta}\partial_{\alpha}({\cal P}^{\alpha\beta}{+}\pit^{\alpha\beta})\right]\Big/({\cal E}{+}{\cal P}_\perp)\;.
\end{eqnarray}
Here $\sigma_{\mu\nu}\equiv\nabla_{\langle\mu}u_{\nu\rangle}$ is the velocity shear tensor and $\theta\equiv\partial{\cdot}u$ the scalar expansion rate.

While the scalar and vector irreducible moments, $\rho_{r}$ and $\rho^{\mu}_{r}$, control the bulk viscous pressure and heat current (which vanish here), we assume that they will not play important roles in the shear dynamics. We will also ignore contributions from irreducible moments of rank ${\,>\,}2$ since they are of higher order in the power-counting scheme discussed below \cite{Denicol:2012cn}. The dynamical equations of motion for $\rhot^{\mu\nu}_{r}$ are obtained \cite{Denicol:2010xn,Denicol:2012cn} by applying the operator $\Delta^{\mu\nu}_{\alpha\beta}D$, where $D\equiv u{\cdot}\partial$ is the time derivative in the LRF, to its kinetic definition (\ref{rhoDef}) and using the Boltzmann equation (\ref{BE}) in the form ($\dot f \equiv Df$)
\begin{equation}
\label{BE2}
\dot{\dft}_\bk=-\dot{f}_{\rs,\bk}-\frac{1}{k{\cdot}u}
       \left(k{\cdot}\nabla\bigl(f_{\rs,\bk}{+}\dft_\bk\bigr)-C_\bk[f]\right).
\end{equation}
Linearizing the collision term in the deviation $\phi_\bk$ one finds \cite{Denicol:2010xn, bazowIP}
\begin{widetext}
\vspace*{-5mm}
\begin{eqnarray}
\label{exactEOM}
    \Delta^{\mu\nu}_{\alpha\beta}\dot{\rhot}^{\alpha\beta}_{r}
    +\sum_{n=0}^{N_2}({\cal A}^{(2)}_{rn})^{\mu\nu}_{\alpha\beta}\rhot^{\alpha\beta}_{n}
    &=& {\cal L}^{\mu\nu}_{r}\dot{\xi} + {\cal M}^{\mu\nu\lambda}_{r}\dot{z}_{\lambda}   
    + \bigl(\alpha^{(2)}_{\theta r}\bigr)^{\mu\nu}\theta
    + \bigl(\alpha^{(2)}_{\sigma r}\bigr)^{\mu\nu\lambda\rho}\sigma_{\lambda\rho}
    + \bigl(\alpha^{(2)}_{\omega r}\bigr)^{\mu\nu\lambda\rho}\omega_{\lambda\rho}
\nonumber\\
    &&+ \frac{{\cal B}^{\mu\nu}_{r}}{\J_{2,0,-1}} \pit^{\alpha\beta}\sigma_{\alpha\beta}
    - \frac{2}{7} (2r{+}5) \rhot^{\lambda\langle\mu}_{r}\sigma^{\nu\rangle}_{\lambda}
    + 2\rhot^{\lambda\langle\mu}_{r}\omega^{\nu\rangle}_{\ \lambda}
    - \frac{1}{3}(r{+}4)\rhot^{\mu\nu}_{r}\theta .
\end{eqnarray}
%
Here  $\omega_{\mu\nu}\equiv(\nabla_{\mu}u_{\nu}-\nabla_{\nu}u_{\mu})/2$ is the vorticity tensor. The tensors $ {\cal L}$, $ {\cal M}$, ${\cal B}$, $\alpha_\theta^{(2)}$, $\alpha_\sigma^{(2)}$, and $\alpha_\omega^{(2)}$ are built from $u^\mu$, $z^\mu$ and $\Delta^{\mu\nu}$, multiplied by thermodynamic integrals over the anisotropic leading-order distribution $f_{\rs,\bk}$ that depend on $\xi$ and $\beta_\rs$; their explicit expressions will be given in \cite{bazowIP}. Our notation follows the convention used in \cite{Denicol:2012cn} as closely as possible. In (\ref{exactEOM}) all comoving time derivatives of $\beta_\rs$ and $u^{\mu}$ were eliminated in terms of spatial gradients and comoving time derivatives of $\xi$ by using Eqs.~(\ref{consLaws}). The matrix ${\cal A}^{(2)}$, which couples moments of different orders $r$, embodies all the microscopic information contained in the linearized Boltzmann collision term:
%
\begin{align}
\bigl({\cal A}^{(2)}_{rn}\bigr)^{\langle \mu \nu\rangle}_{\langle \alpha\beta\rangle }
& =\frac{1}{2}\int dK dK' dP dP'\, W_{\bk\bk'\to\bp\bp'} \,
f_{\rs,\bk}f_{\rs,\bk'} \tf_{\rs,\bp}\tf_{\rs,\bp'} 
(u{\cdot}k)^{r-1}k^{\langle\mu}k^{\nu\rangle} 
\notag \\
& \times \left[ \bigl(\tilde{\cal H}^{(2,2)}_{\bp n}\bigr)^{\langle\lambda\sigma\rangle}%
_{\langle\alpha\beta\rangle} p_{\langle\lambda} p_{\sigma\rangle}
  + \bigl(\tilde{\cal H}^{(2,2)}_{\bp' n})^{\langle\lambda\sigma\rangle}%
_{\langle\alpha\beta\rangle} p'_{\langle\lambda} p'_{\sigma\rangle}
  - \bigl(\tilde{\cal H}^{(2,2)}_{\bk n})^{\langle\lambda\sigma\rangle}%
_{\langle\alpha\beta\rangle} k_{\langle\lambda} k_{\sigma\rangle}
  - \bigl(\tilde{\cal H}^{(2,2)}_{\bk' n})^{\langle\lambda\sigma\rangle}%
_{\langle\alpha\beta\rangle} k'_{\langle\lambda} k'_{\sigma\rangle} \right] .  
\label{linColInt}
\end{align}
\end{widetext}
\vspace*{-9mm}

The main step to close this system of equations is to identify and separate the microscopic time scales. This is accomplished by determining the eigenmodes of ${\cal A}^{(2)}$. Its dimension, which determines the number of eigenmodes, is given by the cutoff $N_\ell$ in the order of the polynomials $P_n^{(\ell)}$ in the expansion (\ref{phi}) of the deviation $\phi_\bk$ which controls the accuracy with which the dependence of $\phi_\bk$ on the LRF energy can be resolved. For $N_\ell\to\infty$ one obtains an infinite number of eigenmodes and associated microscopic time scales.

Introducing the matrix $\Omega^{(2)}$ that diagonalizes ${\cal A}^{(2)}$,
\begin{equation}
\bigl(\bigl(\Omega^{-1}\bigr)^{(2)}\bigr)^{\mu\nu}_{\lambda\rho}\,
\bigl({\cal A}^{(2)}\bigr)^{\lambda\rho}_{\gamma\sigma}\,
\bigl(\Omega^{(2)}\bigr)^{\gamma\sigma}_{\alpha\beta}
=\left[\mathrm{diag}\Bigl(\chi^{(2)}_{0},\dots,\chi^{(2)}_{N_2}\Bigr)\right]^{\mu\nu}_{\alpha\beta},
\notag
\end{equation}
multiplying (\ref{exactEOM}) by $\bigl(\bigl(\Omega^{-1}\bigr)^{(2)}_{ir}\bigr)^{\mu\nu}_{\alpha\beta}$, and summing over $r$ yields
\begin{eqnarray}
\label{Xr}
&&\!\!\!\!\!\!
\dot{X}^{\langle\mu\nu\rangle}_{i}
+(\chi^{(2)}_{i})^{\mu\nu}_{\alpha\beta}X^{\alpha\beta}_{i}
=\bar{\cal L}^{\mu\nu}_{i}\dot{\xi}
{+}\bar{\cal M}^{\mu\nu\lambda}_{i}\dot{z}_{\lambda}
{+}(\bar{\alpha}^{(2)}_{\theta i})^{\mu\nu}\theta\ \ 
\\\notag
&&\!\!\!\!
+(\bar{\alpha}^{(2)}_{\sigma i})^{\mu\nu\lambda\rho}\sigma_{\lambda\rho}
+(\bar{\alpha}^{(2)}_{\omega i})^{\mu\nu\lambda\rho}\omega_{\lambda\rho}
+\mbox{higher-order terms},\quad
\end{eqnarray}
where $X^{\mu\nu}_{r}{\,\equiv\,}\sum_{j=0}^{N_2} \bigl(\bigl(\Omega^{-1}\bigr)^{(2)}_{rj}\bigr)^{\mu\nu}_{\alpha\beta}\,\rhot^{\alpha\beta}_{j}$ are the eigenmodes, and $\bar{\psi}^{\mu\nu\lambda_1\cdots\lambda_n}_{r}\equiv \sum_{j=0}^{N_2}\bigl(\bigl(\Omega^{-1}\bigr)^{(2)}_{rj}\bigr)^{\mu\nu}_{\alpha\beta}\,\psi^{\alpha\beta\lambda_1\cdots\lambda_n}_{j}$ for $\psi \in \{{\cal L}\,,{\cal M}\,,\alpha_\theta^{(2)}\!,\,\alpha_\sigma^{(2)}\!,\,\alpha_\omega^{(2)}\}$. ``Higher-order terms" arise from those proportional to $\rhot_r^{\mu\nu}$ on the r.h.s. of Eq.~(\ref{exactEOM}).

Equation~(\ref{Xr}) shows that the elements of the inverses of the ``eigenvalue tensors'' $[\chi_r^{(2)}]^{\mu\nu}_{\alpha\beta}$ can be interpreted as microscopic relaxation times. Multiplying (\ref{Xr}) by the inverse tensor $[(\chi^{-1})^{(2)}_r]^{\mu\nu}_{\alpha\beta}$ and taking it to zero while keeping $[(\chi^{-1})^{(2)}_r]^{\mu\nu}_{\alpha\beta}\,\bar{\cal L}^{\alpha\beta}_r$ etc. fixed, we see that the term involving $\dot{X}^{\langle\mu\nu\rangle}_{i}$ is also driven to zero and that the eigenmode assumes its ``Navier-Stokes" limit:
\begin{eqnarray}
\label{XrAsy}
&&X^{\mu\nu}_{r}\simeq[(\chi^{-1})^{(2)}_r]^{\mu\nu}_{\alpha\beta}\bar{\cal L}^{\alpha\beta}_{r}\dot{\xi}
+[(\chi^{-1})^{(2)}_r]^{\mu\nu}_{\alpha\beta}\bar{\cal M}^{\alpha\beta\lambda}_{r}\dot{z}_{\lambda}
\\\notag
&&\quad
+[(\chi^{-1})^{(2)}_r]^{\mu\nu}_{\alpha\beta}(\bar{\alpha}^{(2)}_{\theta r})^{\alpha\beta}\theta
+[(\chi^{-1})^{(2)}_r]^{\mu\nu}_{\alpha\beta}(\bar{\alpha}^{(2)}_{\sigma r})^{\alpha\beta\lambda\rho}\sigma_{\lambda\rho}
\\\notag
&&\quad
+[(\chi^{-1})^{(2)}_r]^{\mu\nu}_{\alpha\beta}(\bar{\alpha}^{(2)}_{\omega r})^{\alpha\beta\lambda\rho}\omega_{\lambda\rho}
+\mbox{higher-order terms}.
\end{eqnarray}

The terms explicitly listed on the r.h.s. of Eq.~(\ref{XrAsy}) are proportional to the Knudsen number Kn; the unlisted ``higher-order terms'' all involve an additional factor proportional to one of the modified inverse Reynolds numbers $\sim\rhot_r^{\mu\nu}$ \cite{pc}. 

Next we order the elements of the eigenvalue tensors  $[\chi^{(2)}_r]^{\mu\nu}_{\alpha\beta}$ by magnitude such that $[\chi_{r}^{(2)}]^{\mu\nu}_{\alpha\beta} < [\chi^{(2)}_{r+1}]^{\mu\nu}_{\alpha\beta}$. We consider the transient regime when the slowest mode ($r\eq0$) still evolves dynamically according to Eq.~(\ref{Xr}) while all faster modes with $r{\,\geq\,}1$ have already reached their asymptotic Navier-Stokes limits (\ref{XrAsy}). Inverting the relation between eigenmodes and irreducible moments, $\rhot^{\mu\nu}_{i}=\sum_{j=0}^{N_2}(\Omega^{(2)}_{ij})^{\mu\nu}_{\alpha\beta}X^{\alpha\beta}_{j}$, normalizing the diagonalizing matrix by requiring $(\Omega^{(2)}_{00})^{\mu\nu}_{\alpha\beta}= \mathbb{1} \cdot \Delta^{\mu\nu}_{\alpha\beta}$, and remembering $\pit^{\mu\nu}\eq\rhot^{\mu\nu}_0$, Eq.~(\ref{XrAsy}) yields
\begin{eqnarray}
\label{rho}
&&\!\!\!\!
\rhot^{\mu\nu}_{i}= (\Omega^{(2)}_{i0})^{\mu\nu}_{\alpha\beta}\pit^{\alpha\beta}+\hat{\ell}^{\mu\nu}_{i}\dot{\xi}
+\hat{m}^{\mu\nu\lambda}_{i}\dot{z}_{\lambda}
+(\hat{\eta}^{(2)}_{\theta i})^{\mu\nu}\theta
\nonumber\\
&&\hspace{4mm}
+\,(\hat{\eta}^{(2)}_{\sigma i})^{\mu\nu\lambda\rho}\sigma_{\lambda\rho}
+(\hat{\eta}^{(2)}_{\omega i})^{\mu\nu\lambda\rho}\omega_{\lambda\rho} 
+\text{h.-o.\,terms.}
\qquad
\end{eqnarray}
Here we defined $\hat \ell$ in terms of ${\cal L}$ by
\beq
\hat{\ell}^{\mu\nu}_{i}\equiv\ell^{\mu\nu}_{i}-(\Omega^{(2)}_{i0})^{\mu\nu}_{\alpha\beta}\ell^{\alpha\beta}_{0},\ \ 
\ell^{\mu\nu}_{i}\equiv\sum_{r=0}^{N_2}(\tau^{(2)}_{ir})^{\mu\nu}_{\alpha\beta}{\cal L}^{\alpha\beta}_{r}, \ \ 
\eeq
and similarly for $\hat{m}$, $\hat\eta_\theta^{(2)}$, $\hat\eta_\sigma^{(2)}$ and $\hat\eta_\sigma^{(2)}$ in terms of ${\cal M}$, $\alpha_\theta^{(2)}$, $\alpha_\sigma^{(2)}$ and $\alpha_\omega^{(2)}$, with the microscopic relaxation times
\beq
(\tau^{(2)}_{in})^{\mu\nu}_{\alpha\beta}\equiv
\sum_{m=0}^{N_2}(\Omega^{(2)}_{im})^{\mu\nu}_{\lambda\rho}
[(\chi^{-1})^{(2)}_{m}]^{\lambda\rho}_{\gamma\sigma}
\bigl[(\Omega^{-1})^{(2)}_{mn}\bigr]^{\gamma\sigma}_{\alpha\beta}.
\eeq
Equation~(\ref{rho}), built on Eq.~(\ref{XrAsy}), implements the Navier-Stokes limit for all microscopic eigenmodes except the slowest one, $r\eq0$. It expresses all higher-order tensor moments in terms of the lowest-order one, $\rhot^{\mu\nu}_0\eq\pit^{\mu\nu}$, which appears in the energy-momentum conservation law. This is the key step in closing the  equation of motion (\ref{exactEOM}) to simplify the coupling to higher order irreducible moments. To this end we multiply Eq.~(\ref{exactEOM}) by $(\tau^{(2)}_{nr})^{\mu \nu}_{\alpha\beta} = [({\cal A}^{(2)})^{-1}_{nr}]^{\mu\nu}_{\alpha\beta}$ and insert Eq.~(\ref{rho}). Some algebra yields a set of equations, labeled by the order $n$ of the irreducible moment considered, of the following form:
\begin{eqnarray}
\label{EOM}
\!\!\!\!\!\!\!\!
&&(\tau^{(n)}_\pi)^{\mu\nu}_{\alpha\beta}\dot{\pit}^{\alpha\beta}
+(\Omega^{(2)}_{n0})^{\mu\nu}_{\alpha\beta}\pit^{\alpha\beta}
+(\tau^{(n)}_\xi)^{\mu\nu}\dot{\xi}
+(\lambda^{(n)}_z)^{\mu\nu\lambda}\dot{z}_\lambda
\notag\\
&&\quad
+ ({\cal D}^{(n)}_2)^{\mu\nu}+({\cal D}^{(n)}_1)^{\mu\nu}
\nonumber\\
&&=(\eta_\theta^{(2,n)})^{\mu\nu}\theta
     +(\eta_\sigma^{(2,n)})^{\mu\nu\alpha\beta}\sigma_{\alpha\beta}
     +(\eta_\omega^{(2,n)})^{\mu\nu\alpha\beta}\omega_{\alpha\beta}
\nonumber\\
&&\quad+({\cal J}^{(n)})^{\mu\nu}+({\cal K}^{(n)})^{\mu\nu}\;.
\end{eqnarray}
The coefficients of the various terms are related to (as indicated by the chosen notation) but not identical with those in Eq.~(\ref{rho}); their explicit expressions will be given in \cite{bazowIP}. The first two terms on the left and the terms on the right of Eq.~(\ref{EOM}) are similar in structure to those of the generalized Israel-Stewart theory (DNMR theory) derived in \cite{Denicol:2012cn}; the more complex Lorentz tensor structure results from the local momentum anisotropy of our leading-order distribution function (\ref{fa}). As in \cite{Denicol:2012cn}, the ${\cal J}$ and ${\cal K}$ terms are ${\cal O}(\mathrm{Kn}^2)$
and ${\cal O}(\mathrm{Kn}\cdot\tilde{\mathrm{R}}^{-1}_{i})$, respectively \cite{fn3}. The next two terms on the left of (\ref{EOM}) arise from the second and third term on the right of Eq.~(\ref{rho}), originating in the momentum anisotropy of the leading-order distribution $f_{\rs,\bk}$. The last two terms on the left of Eq,~(\ref{EOM}) originate from time derivatives of these terms when evaluating the first term in Eq.~(\ref{exactEOM}). They include (in $({\cal D}^{(n)}_{2})^{\mu\nu}$) second time derivatives $\sim\ddot{\xi},\,\ddot{z}^{\lambda}$ of the parameters $\xi,\,z^\lambda$ characterizing the deviation from local momentum isotropy in $f_{\rs,\bk}$, as well as products of first time derivatives of these parameters with other first derivatives (${\sim\,}\dot{z}{\cdot}\dot{z},\,  \dot{z}^{\lambda}\dot{z}^{\rho},\,  \dot{\xi}\dot{z}^{\lambda},\, \dot{\xi}{\cdot}{\cal O}(\mathrm{Kn},\tilde{\mathrm{R}}^{-1}_{i}),\, \dot{z}^{\lambda}{\cdot}{\cal O}(\mathrm{Kn},\tilde{\mathrm{R}}^{-1}_{i})$) (these are collected in $({\cal D}^{(n)}_{1})^{\mu\nu}$). Such terms occur generically whenever the Boltzmann equation is expanded around a leading-order distribution function that is locally anisotropic in momentum space \cite{fn4}. As we will see below, the terms involving second time derivatives of $\xi$ identify this anisotropy parameter as a quasi-normal mode of the non-equilibrium dynamics which undergoes transient oscillations that are damped on a microscopic time scale that is related to the elements of $(\tau^{(n)}_\xi) ^{\mu\nu}$ and a similar coefficient multiplying the $\ddot{\xi}$ term in $({\cal D}^{(n)}_{2})^{\mu\nu}$ \cite{fn5}. Such quasi-normal modes appear to be a generic feature in strongly-coupled theories which cannot be decribed by the Boltzmann equation \cite{Starinets:2002br, Nunez:2003eq, Noronha:2011fi,Chesler:2008hg,Heller:2014wfa
}. 

For $n\eq0$, Eq.~(\ref{EOM}) controls the dynamical evolution of the residual shear stress $\pit^{\mu\nu}$. It is coupled to the dynamical evolution of $\xi$ for which we can take Eq.~(\ref{EOM}) for $n\eq1$. (When expanding around an isotropic leading-order distribution this equation is not needed.) Eqs.~(\ref{EOM}) for $n\eq1$, together with its siblings for the bulk viscous pressure and heat flow vector, are sufficient to solve for all possible momentum-space deformation parameters of the leading-order distribution (9 at most). For $n{\,>\,}1$, Eqs.~(\ref{EOM}) are only needed if we want to include dynamics at even shorter time scales, by making $\rhot_{j>0}^{\mu\nu\lambda\cdots}$ dynamical variables instead of using their Navier-Stokes limits (\ref{XrAsy}) as we have done here.


\section {Eigenmode analysis} 
\begin{figure}[t]
\includegraphics[width=0.8\linewidth]{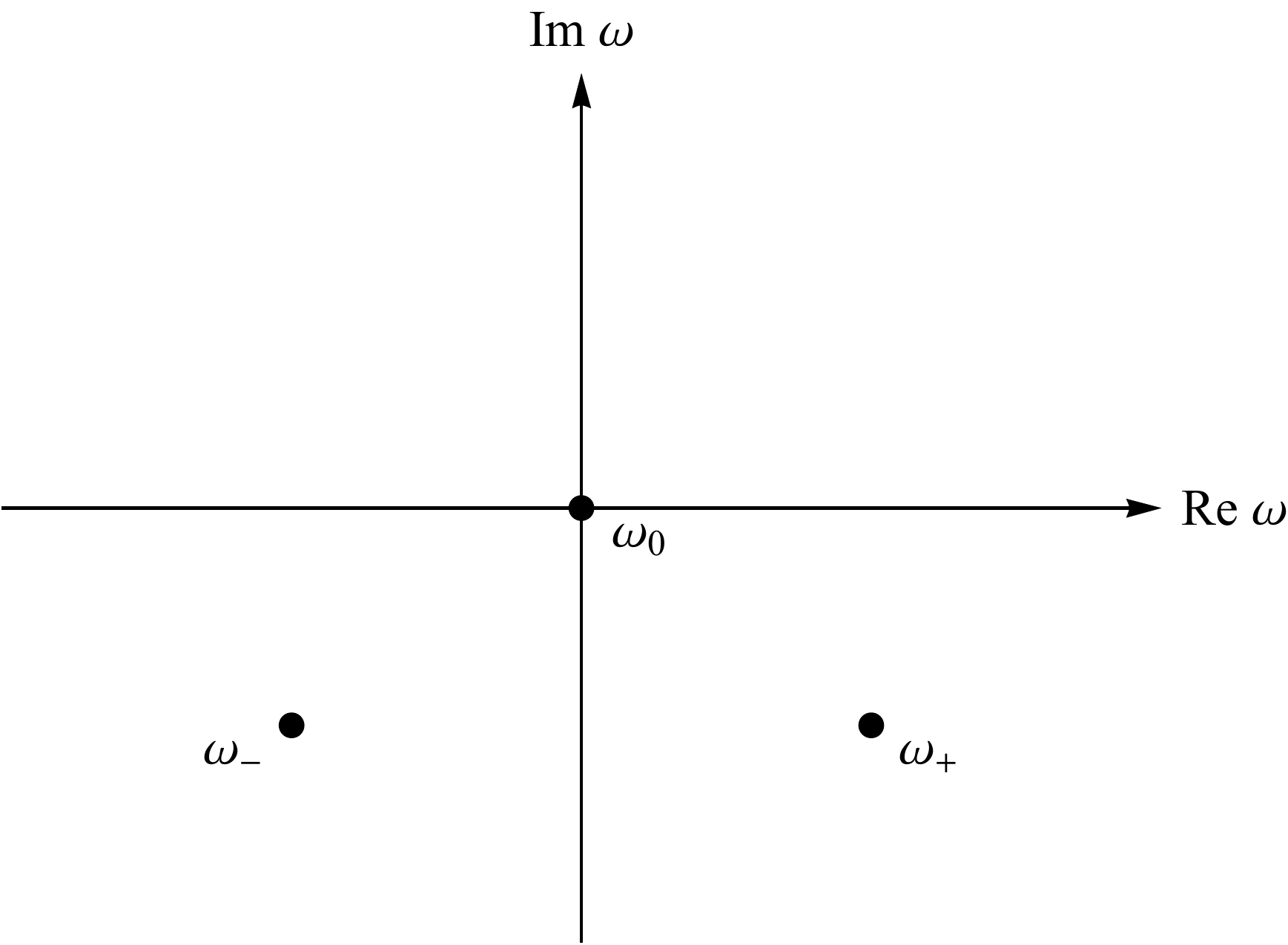}
\vspace{-2mm}
\caption{Shear channel eigenmodes in the zero wavenumber limit. Shown are the two degenerate hydrodynamic modes at $\omega_0\eq0$ and the first two non-hydrodynamic modes at $\omega_\pm$ that exhibit transient oscillatory behavior.}
\label{fig:1}
\end{figure}

To obtain dispersion relations for the eigenmodes of the linearized equations of motion, we consider a linear perturbation around a static background~\cite{DeGroot}: 
\begin{eqnarray}
\xi&=&\xi_0+ e^{i\omega t-ikz}\,\delta\xi \;,
\nonumber\\\nonumber
\pit^{\mu\nu}&=&\pit^{\mu\nu}_0+e^{i\omega t-ikz}\,\delta\pit^{\mu\nu} \;,
\\
u^{\mu}&=&u^{\mu}_0+e^{i\omega t-ikz}\,\delta u^{\mu} \;,
\end{eqnarray}
with $\xi_0\eq\mathrm{const.}{\,\ne\,}0$, $\pit^{\mu\nu}_{0}\eq0$, and $u^{\mu}_0=(1,0,0,0)$, respectively. To study the shear channel we consider flow in the $y$-direction with a gradient in the $z$-direction, $\delta u^{\mu}=(0,0,\delta u^{y}(t,z),0)$, and study $\pit^{yz}$. For simplicity, we assume $\delta z^{\mu}\eq0$, no vorticity, and constant temperature; the system is then not expanding, i.e. $\theta\eq0$. In the linear regime, the elements of $({\cal D}^{(n)}_1)^{\mu\nu}$, $({\cal J}^{(n)})^{\mu\nu}$, $({\cal K}^{(n)})^{\mu\nu}$ are all terms of ${\cal O}(\delta^2)$ and can therefore be ignored \cite{fn6}. In $({\cal D}^{(n)}_2)^{\mu\nu}$ only one term $\sim\ddot{\xi}$ contributes \cite{bazowIP} which we here denote as $(\tilde{\xi}^{(n)})^{\mu\nu}\ddot{\xi}$. Introducing the shorthand notations
\begin{eqnarray}
&\eta^{(n)}\equiv(\eta_\sigma^{(2,n)})^{yzyz}\;,\quad
&\tilde{\xi}^{(n)}\equiv(\tilde{\xi}^{(n)})^{yz}\;,
\nonumber\\
&\tau^{(n)}_{\pi}\equiv(\tau^{(n)}_{\pi})^{yz}_{yz}\;,\quad
&\tau^{(n)}_{\xi}\equiv(\tau^{(n)}_{\xi})^{yz}\;,
\nonumber\\
&\Omega_{n}\equiv(\Omega^{(2)}_{n0})^{yz}_{yz}\;,\quad
&d\equiv\partial_{\xi}{\cal P}^{yz}\;,
\end{eqnarray}
the linearized equations of motion can then be written as $A\bm{x}\eq0$, with
\begin{eqnarray}
\label{eq:mateq}
&&A\equiv \begin{pmatrix}
  i\omega({\cal E}+{\cal P}_\perp) & -ikd & -ik  \\
  -ik\eta^{(0)} & -\omega^2\tilde{\xi}^{(0)}+i\omega\tau^{(0)}_\xi & i\omega\tau^{(0)}_\pi+\Omega_{0} \\
  -ik\eta^{(1)} & -\omega^2\tilde{\xi}^{(1)}+i\omega\tau^{(1)}_\xi & i\omega\tau^{(1)}_\pi+\Omega_{1} \\
 \end{pmatrix}\;,
\nonumber\\
&&\bm{x}\equiv \begin{pmatrix}
  \delta u^{y}, & \delta\xi, & \delta\pit^{yz} 
 \end{pmatrix}^T\;.
\end{eqnarray}
We note that the coefficients $\tilde{\xi}^{(n)}$ vanish when the expansion (\ref{momentexpansion}) is truncated at $N_2\eq0$, so in Eq.~(\ref{exactEOM}) $N_2$ must be $\geq1$. For nontrivial solutions 
of (\ref{eq:mateq}) the determinant of $A$ must vanish. In the long wavelength limit $k\to0$ this determinant reduces to a fourth-order polynomial in $\omega(0)\eq\lim_{k\to0}\omega(k)$, with two degenerate zeros  at $\omega_0\eq0$ describing hydrodynamic modes, and two non-hydrodynamic modes corresponding to solutions of the quadratic equation
\begin{eqnarray}
\label{w}
&& \omega^2\left(\tau^{(1)}_\pi \tilde{\xi}^{(0)}{-}\tau^{(0)}_\pi\tilde{\xi}^{(1)}\right)
\nonumber\\
&&\ +\,i\omega\left(\tau^{(0)}_\pi\tau^{(1)}_\xi{-}\tau^{(1)}_\pi\tau^{(0)}_\xi{+}\Omega_{0}
                      \tilde{\xi}^{(1)}{-}\Omega_{1}\tilde{\xi}^{(0)}\right)
\nonumber\\
&&\ +\left(\Omega_{0}\tau_\xi^{(1)}{-}\Omega_{1}\tau_\xi^{(0)}\right)=0
\end{eqnarray}
that have the form $\omega_{\pm}\equiv\pm\omega_r{-}i\omega_i$, with $\omega_r,\ \omega_i>0$ (see 
Fig.~\ref{fig:1}). [That $\omega_i>0$ is not obvious from (\ref{w}) but necessary for the system to dynamically evolve towards local momentum isotropy.] The equal and opposite real and the negative imaginary parts of the frequencies of the non-hydrodynamic modes indicate that the momentum anisotropy parameter $\xi$ undergoes transient (damped) oscillations with frequency $\omega_r$ and damping rate $\omega_i$. Similar transient oscillations were found in strongly coupled plasmas when described using the fluid/gravity correspondence \cite{Starinets:2002br, Nunez:2003eq, Chesler:2008hg, Noronha:2011fi, 
Heller:2014wfa}. 

\section {Conclusions} In strongly coupled systems it was previously shown \cite{Heller:2014wfa} that the transition between pre-equilibrium dynamics at early and hydrodynamic behavior at later stages of a nuclear collision is characterized by transient oscillations of the slowest non-hydrodynamic modes, with (overdamped) oscillation amplitudes decaying on microscopic time scales. The present work, which is based on weakly coupled microscopic dynamics using the Boltzmann equation, suggests that these oscillations arise from large deviations from local momentum isotropy in rapidly and anisotropically expanding systems: When strongly anisotropic expansion drives the local momentum distribution away from isotropy, the parameters describing this anisotropy do not decay exponentially, but perform damped oscillations. The matrix $A$ in Eq.~(\ref{eq:mateq}) couples their oscillations to $\pit^{\mu\nu}$ which therefore oscillates in sync. These transient oscillations are a generic phenomenon, but capturing them in the macroscopic equations of motion requires the inclusion of expansion-driven local momentum anisotropies in the microscopic distribution function already at leading order. If the distribution function is instead expanded around a locally isotropic equilibrium state when deriving the macroscopic equations of motion from moments of the Boltzmann equation, only exponentially damped non-hydrodynamic modes are found \cite{Denicol:2011fa}. 


\acknowledgments
We thank J.~Noronha and M.~Heller for useful discussions. U.H. thanks the Institute for Nuclear Theory for hospitality during the final stages of this work. Financial support by the U.S. Department  of Energy, Office of Science, Office of Nuclear Physics under Awards No. \rm{DE-SC0004286} and (within the framework of the JET Collaboration) \rm{DE-SC0004104} is gratefully acknowledged.



\begin{thebibliography}{99}
\bibitem{fn1}
For anisotropic collective expansion with a constant expansion rate the distribution function would approach this quasi-equilibrium form, with an appropriately chosen set of anisotropy parameters.
\bibitem{fn2a}
  Note that no symmetry assumptions are made for the transition rate $W_{\bk\bk'\to \bp\bp'}$.
\bibitem{fn2} 
  To reduce clutter we write $\rhot^{\mu_1\cdots\mu_\ell}_{r}(x)$ instead of 
  $\rhot^{\langle\mu_1\cdots\mu_\ell\rangle}_{r}(x)$, remembering that the moments
  are traceless and have only spatial components in the LRF.
\bibitem{fn3}
   Terms of second order in the inverse Reynolds number are absent in 
   Eq.~(\ref{EOM}) because we only kept terms linear in the deviation 
   $\phi_\bk$ when expanding the collision integral around the leading-order 
   distribution.
\bibitem{fn4}
Equation (\ref{fa}) is just one such example; our results are generic and apply for any leading-order distribution with $x$-dependent anisotropy parameters, irrespective of its specific functional form.
\bibitem{fn6}
  Some terms in $({\cal K}^{(n)})_{\mu\nu}$ would render the constitutive equations 
  acausal and thus need to be discarded. This happens automatically in the linearized
  approximatiion studied here.
\end{thebibliography}
\end{document}